# Gallium-doped Zinc Oxide: Nonlinear Reflection and Transmission Measurements and Modeling in the ENZ Region

*Adam Ball[1*], Ray Secondo[1,2], Benjamin T. Diroll[3], Dhruv Fomra[1], Kai Ding[1], Vitaly Avrutin[1], Ümit Özgür[1], and Nathaniel Kinsey[1]*

1. Department of Electrical and Computer Engineering, Virginia Commonwealth University, Richmond, VA, 23284, USA
2. Air Force Research Laboratory Wright-Patterson Air Force Base Dayton, OH, 45433, USA
3. Center for Nanoscale Materials, Argonne National Laboratory, 9700 S. Cass Avenue, Lemont, IL, 60439, USA
E-mail: ballar@vcu.edu

**Abstract (200-word limit)**

Strong nonlinear materials have been sought after for decades for applications in telecommunications, sensing, and quantum optics. Gallium-doped zinc oxide is a II-VI transparent conducting oxide that shows promising nonlinearities similar to indium tin oxide and aluminum-doped zinc oxide for the telecommunications band. Here we explore its nonlinearities in the epsilon near zero (ENZ) region and show $n_{2,eff}$ values on the order of $4.5\times10^{-3}$ cm$^2$GW$^{-1}$ for IR pumping on 200-300 nm thin films. Measuring nonlinear changes in transmission and reflection with a white light source probe in the near-IR while exciting in the near-IR provides data in both time and wavelength. Three films varying in thickness, optical loss, and ENZ crossover wavelength are numerically modeled and compared to experimental data showing agreement for both dispersion and temporal relaxation. In addition, we discuss optimal excitation and probing wavelengths occur around ENZ for thick films but are red-shifted for thin films where our model provides an additional degree of freedom to explore. Obtaining accurate nonlinear measurements is a difficult and time-consuming task where our method in this paper provides experimental and modeled data to the community for an ENZ material of interest.

## 1. Introduction

Nonlinear optics is known to be limited by its weak light-matter interaction[1,2] which has pushed researchers to find improved, robust materials [3,4]. Recently, transparent conducting oxides (TCOs) have become an active contender for nonlinear optics because of their strong nonlinearities [5–7] and ease of their integration with current platforms like silicon photonics.[8–14] The driving component for the nonlinearities is the permittivity stemming from the number of free electrons in the semiconductor. These materials have an epsilon-near-zero (ENZ) region naturally in the telecommunication range (1550 nm) with doping densities ~10$^{20}$ cm$^{-3}$ that can enhance the light-matter interaction due to the slow light effect and D-field continuity. [11–24] Materials such as indium tin oxide (ITO) [15–17] and aluminum-doped zinc oxide (AZO) [18–20] have been popular in the last few years with various enhanced nonlinear properties reported, [21].

For example, ENZ materials show adiabatic frequency shifting ($\Delta f \sim$ THz),[22–29] large amplitude modulation ($\Delta T > 50\%$),[30–32] and extreme changes in refractive index ($\Delta n \sim 1$).[33–35] In addition, when coupled with metastructures, these ENZ films can further enhance the nonlinear properties achieved.[26,35–41] Gallium-doped zinc oxide (GZO) is proposed to show similar nonlinear optical performance to AZO films but has not been thoroughly studied experimentally. GZO has been used as a TCO for solid state lighting,[42,43] photovoltaics,[44–46] shown promising plasmonic properties,[47,48] and is an alternative to AZO. Performing nonlinear experiments are challenging and time consuming, where our experimental method allows us to probe the nonlinear coefficients for many wavelengths simultaneously. When data in literature is limited or restricted, it limits what others can create with it. Here we intend to provide a wide-ranging dataset of GZO linear and nonlinear optical parameters, currently absent from literature.

In this work, GZO films grown by molecular beam epitaxy (see supplemental S1) are studied through pump-probe nonlinear reflection and transmission measurements showing large modulation for multiple IR (intraband) excitation wavelengths. The transients are explored using a white light probe ($\lambda_{probe} = 1000$-$1600$ nm) providing a vast array of data in both time and probe wavelength. Three GZO films are used in this study to explore the intraband nonlinearities where each sample is composed of a GZO thin film (~200 nm), a ZnO buffer layer (~25 nm), a MgO buffer layer (~5 nm), and a c-plane sapphire substrate. **Table 1** describes the three GZO films used in this study labeled as samples 1, 4, and 5 (S1, S4, and S5) accordingly and referenced as so in future sections. Each film has a unique crossover wavelength ($\lambda_{CO}(\text{Re}\{\varepsilon = 0\})$), loss at crossover, GZO thickness, and ZnO thickness. The permittivity and optical properties of each film are extracted from the J. A. Woollam M-2000 ellipsometer and can be found in **Figure 1** (b) and supplemental A2.

**Table 1.** The three Ga:ZnO samples used for nonlinear measurements each with a unique crossover wavelength, loss at crossover wavelength, and thicknesses of GZO and ZnO layers.

| Sample | Crossover λ [nm] | ε'' at $\lambda_{CO}$ | GZO [nm] | ZnO [nm] | $\varepsilon_\infty$ | Carrier Conc. [cm$^{-3}$] |
|---|---|---|---|---|---|---|
| S1 | 1560 | 0.31 | 207 | 21 | 3.5 | 6.07×10$^{20}$ |
| S4 | 1700 | 0.41 | 225 | 24 | 3.7 | 5.28×10$^{20}$ |
| S5 | 1710 | 0.31 | 280 | 25 | 3.7 | 5.17×10$^{20}$ |

In addition to the figures and information provided, a dataset of all the GZO measurements and code is provided in the supplemental information and online within a nonlinear ENZ application.[49] This is a vast dataset of all the samples, wavelengths, and intensities measured whereas the figures in this manuscript only cover a fraction of the data measured. In addition to all the measurements, the models described in R. Secondo et al. are used to fit the GZO nonlinear measurements.[50] This further verifies the results of the measurements and nonlinear coefficients extracted. The script is divided into five sections: the introduction, linear material properties, IR excitation properties including dispersion and temporal data, thickness dependence of ENZ thin films, and concluding remarks.

## 2. Linear GZO Properties

The linear properties of the films are extracted using a transfer-matrix-method code that accounts for all 4 layers of the sample.[51] The GZO layer is modeled by a modified Drude equation given by equation 1[52] while the ZnO, MgO, and sapphire layers are modeled by static refractive indices of 1.95, 1.7, and 1.75, respectively. From ellipsometry data, the dispersion of these layers is minimal in the 1000-1700 nm wavelength range.

$$\varepsilon(\omega) = \varepsilon_\infty - \frac{Nq^2}{\varepsilon_0 m^*} \frac{1}{(\omega^2 - i\Gamma\omega)} \quad (1)$$

Extraction of the nonlinear parameters is done by using equations 2 and 3. The absorption coefficient is calculated by $\Delta\alpha = 4\pi\Delta k/\lambda$, where $\Delta k = \Delta Im\{\sqrt{\varepsilon}\}$.

$$n = n_0 + n_{2,eff}I \rightarrow \Delta n = n_{2,eff}I \quad (2)$$

$$\alpha = \alpha_0 + \beta_{eff}I \rightarrow \Delta\alpha = \beta_{eff}I \quad (3)$$

Although the nonlinear changes of the optically thin ZnO and MgO layers are minor compared to the changes in GZO, the contributions of each layer cannot be separated entirely. A sample of the ZnO, MgO, and sapphire substrate were measured at all three pump wavelengths and exhibited a ~2% $\Delta R/R$ and $\Delta T/T$ contribution to the nonlinearity shown in supplemental A3. These contributions are minimal compared to the GZO layer itself and are within other experimental errors. Additionally, the nonlinearities measured in normalized reflection and transmission are not strictly due to Kerr effects, rather free carrier effects in GZO. Consequentially, in agreement with best nonlinear measurement practices, the nonlinear coefficients reported in this work are presented as effective nonlinear refraction ($n_{2,eff}$) and nonlinear absorption ($\beta_{eff}$).

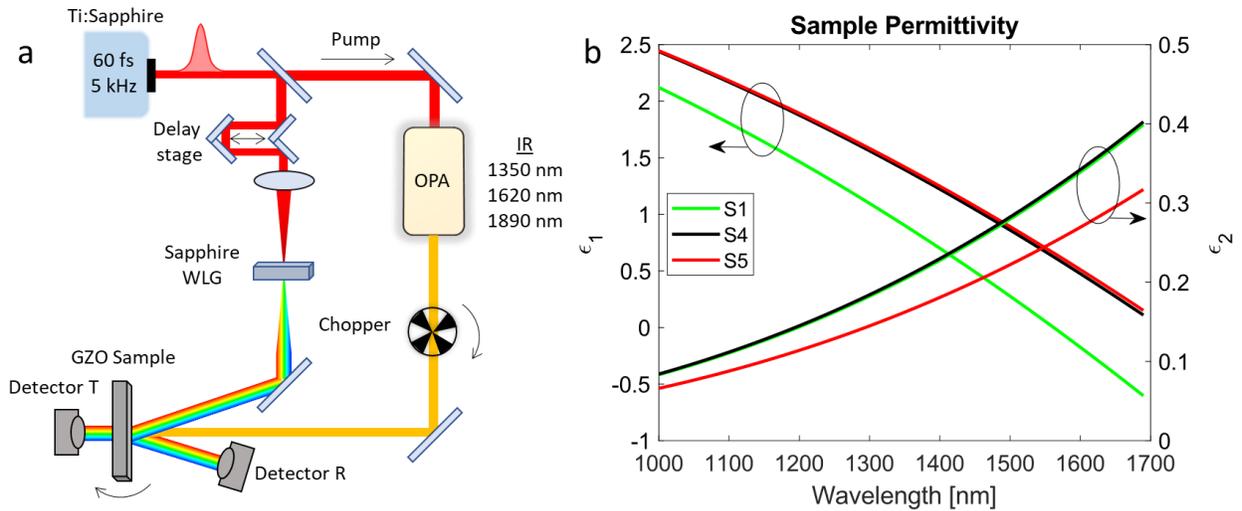

**Figure 1.** a) Experimental setup for $\Delta R/R$ and $\Delta T/T$ nonlinear measurements with a white light probe and IR pump from an OPA. b) Permittivity graphs of S1, S4, and S5 showing both real (left y-axis) and imaginary (right y-axis) components.

## 3. IR Excitation

Free carrier effects dominate the nonlinearities inside degenerately doped GZO. For IR energies, absorption of photons causes conduction band electrons to rapidly jump in higher energy levels *within* the band, hence it is termed *intra*-band nonlinearities.

$$\Delta U(t) = A(\lambda_{IR}, t) \frac{I_{0IR} t_{IR}}{d} \eta_{IR}(t) \qquad (4)$$

Where $A(\lambda_{IR},t)$ is the absorption of IR light, $I_{0IR}$ is the intensity of the light, $t_{IR}$ is the pulse duration, $d$ is the thickness of the film, $\eta_{IR}$ is the temporal response of an error function multiplied with an exponential decay.

This increase in energy (equation 4) causes the electron to move higher in the *E-k* band diagram thus changing the electron's effective mass because of conduction band non-parabolicity.[53] This increase in effective mass pushes the permittivity to longer wavelengths according to the plasma frequency in the Drude equation (1) and subsequently changes the reflection and transmission properties of the film. In this study, pump wavelengths of 1350, 1620, and 1890 nm are used. Considering the crossover wavelength of each of these films is between 1560-1710 nm, the pump wavelengths used to modulate the films in wavelength ranges shorter, around, and longer than the ENZ point to investigate best pumping regimes. Average pump powers of 1.3-21 mW are used to excite the transient transmission and reflection spectra measured by the response of the white light probe.

**Figures 2** (a) and (b) show example dispersion and delay surface plots for reflection and modeled data for S5 pumped at 1890 nm with 8.0 mW power. Subsequently, Figures 2 (c-d) are the nonlinear parameters calculated in both time and wavelength based on the change in transmission and reflection from the IR pump. The peak change in the real index is $\Delta n \sim 0.1$ and the peak change in absorption is $\Delta \alpha \sim -3 \times 10^5$ m$^{-1}$. The overall change in refractive index is on the scale of the linear index and shows the strength of ENZ materials in nonlinear optics. Moreover, the peak in index modulation occurs at 1600 nm, the edge of our measurement range. However, the theory in later sections will show, based on thickness, that the nonlinearity will peak closer to the ENZ point. The zero-crossing point in Figure 2 a and b arises from the red-shift of the reflection curve from its previous position, a point of $\Delta R = 0$ occurs because the reflection curve's positive concavity. At shorter wavelengths, reflection increases; at longer wavelengths reflection decreases.

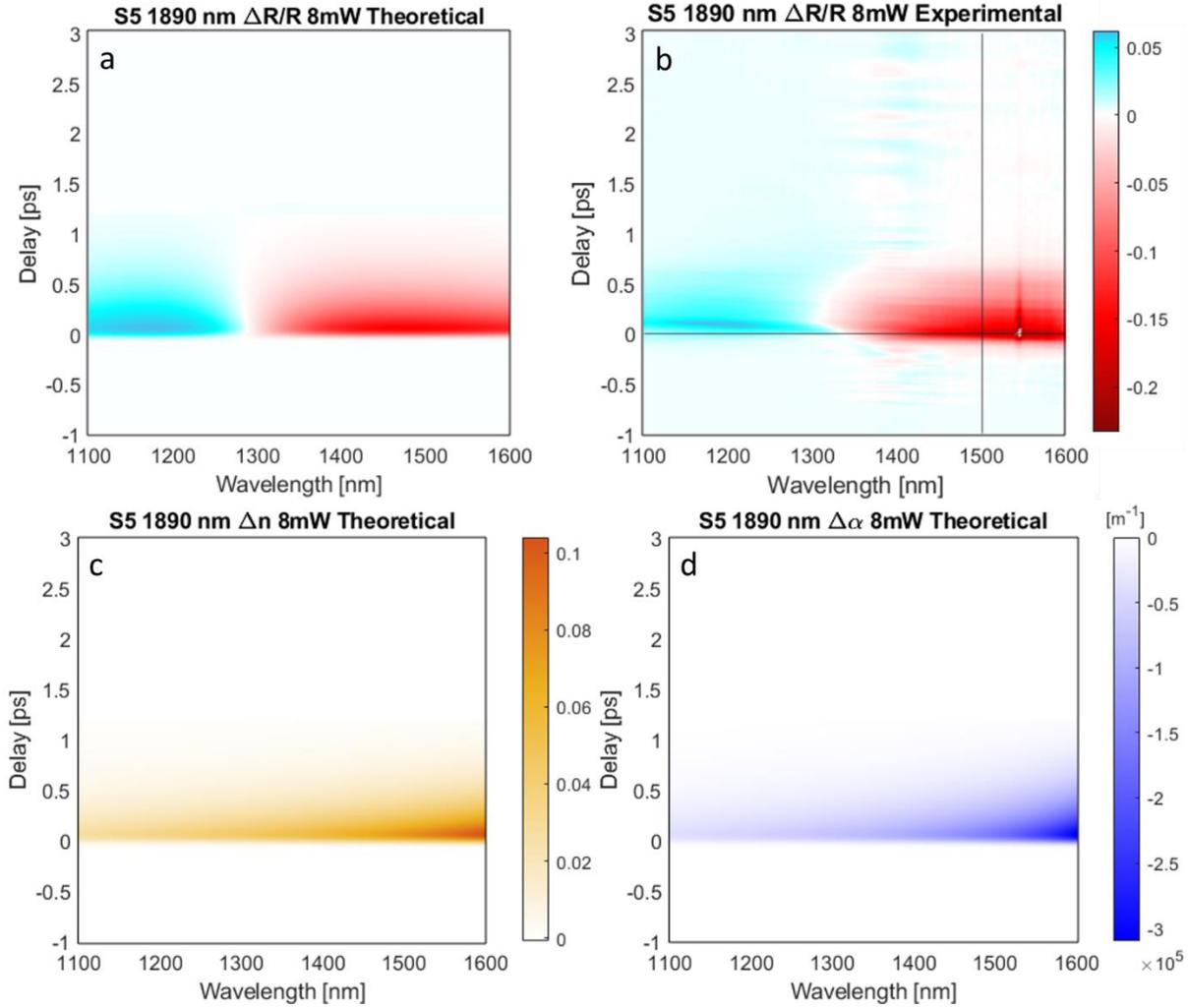

**Figure 2.** a) Calculation of nonlinear reflection change in S5 due to the 1890 nm pump matching plot b). b) Experimental reflection data from S5 excited with 1890 nm with a power of 8.0 mW. Cross sections of t = 0 and $\lambda = 1500$ nm are used for modeling in Figure 3. c) Calculated change in refractive index in S5 due to the pump. d) Numerically calculated change in $\alpha$ in S5 due to the pump.

The cross sections for S5 pumped at 1890 nm and powers varying from 1.3-21 mW, where 8 mW is signified by the horizontal and vertical lines in Figure 2 (b), are shown in **Figure 3**. The wavelength cross section shows all the intensities used on this film and the associated fit numerically modeled from previous works.[50] The delay cross sections in Figure 3 (a) and (c) show the intensities and the modeling of the temporal dynamics. The delay curves are composed of a fast electron response (~10s of fs) and an IR relaxation of $\tau_{IR}$ ~ 300 fs defined by 1/e decay fitted by an exponential model. The purpose of the single exponential decay is to capture the electron nonlinearity of interest. In Figure 3a, there is a decrease in the normalized transmission around t = 1ps, speculated to be a residual counter-acting bulk film interaction unrelated to the hot-electrons because of its long relaxation time which is not currently included within the model. This effect is linear in power and decreases the overall peak change in transmission and thus describes why the model overestimates the nonlinearity at high intensities. The wavelength cross sections in Figure 3 (b) and (d) signify the correct spectral trends of the film for both transmission and reflection thus enabling us to accurately model the change in refractive index.

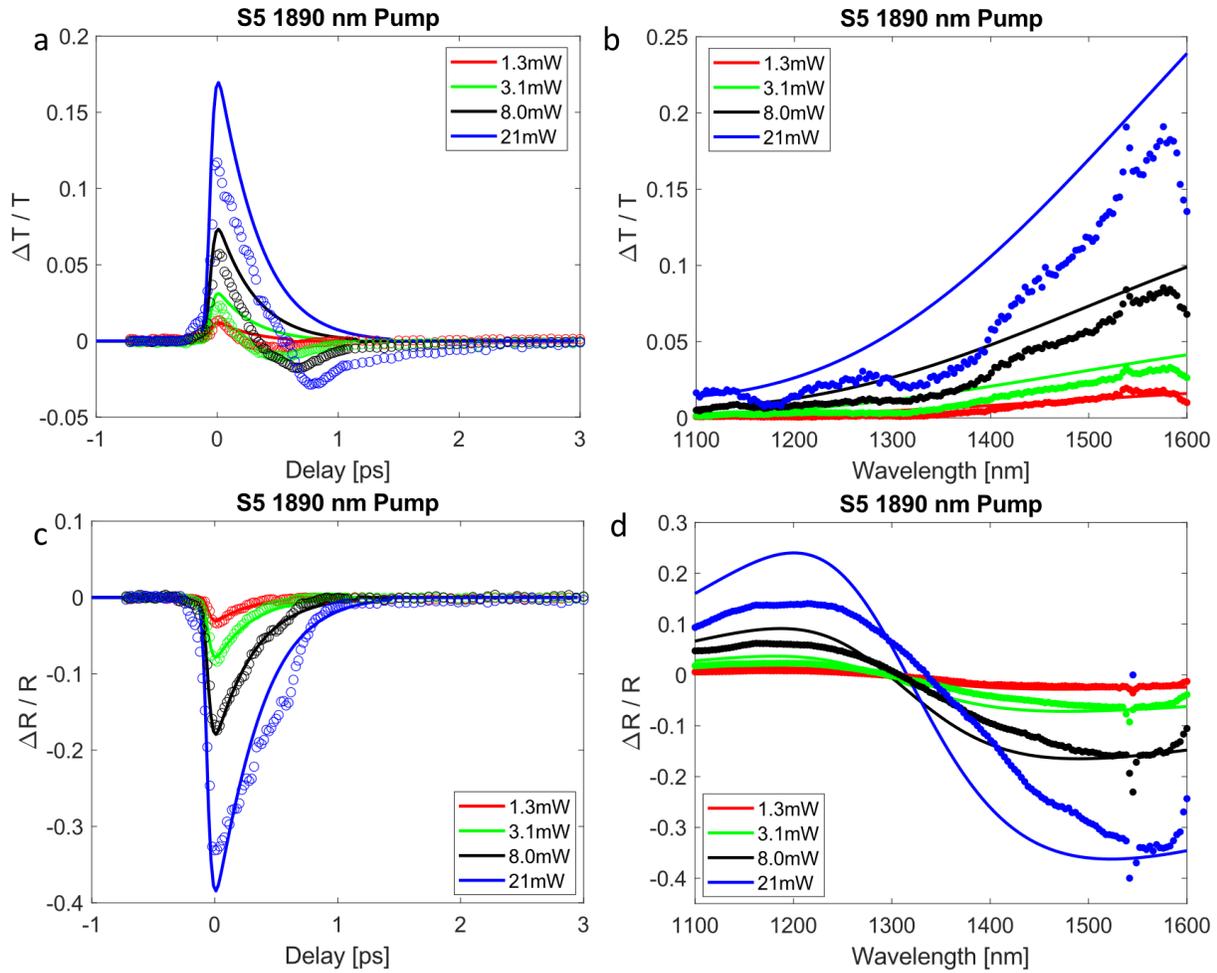

**Figure 3.** a) Cross section of Δ*T/T* temporal dynamics at λ$_{probe}$ = 1500 nm. b) Cross section of Δ*T/T* dispersion dynamics at *t* = 0. c) Cross section of Δ*R/R* temporal dynamics at λ$_{probe}$ = 1500 nm. d) Cross section of Δ*R/R* dispersion dynamics at *t* = 0. The data is S5 pumped at 1890 nm. Each line corresponds with increasing average powers of 1.3 mW, 3.1 mW, 8 mW, and 21 mW. Dots represent data measured in experiments and solid lines represent the theoretical curves.

## 4. Thickness Dependence of ENZ Thin Films

When looking at the nonlinear Kerr effect, one of the most important linear parameters that dictates the nonlinearities is absorption. When working with films of varying thickness (200-300 nm) it is imperative to know the highest linear absorbing region to best facilitate the nonlinear mechanism. In typical materials defined by the Drude model, the dielectric regime (ε' > 1) is a spectral region of high transmission, and the metallic region (ε' < 0) is a spectral region of high reflection. The crossover point between reflection and transmission is where absorption peaks from energy conservation. As the thickness decreases in an ENZ film, the shift between high transmission and high reflection moves further into the IR wavelengths. For a film with the same properties as S1, but with varying thickness, the simulated normalized absorption is plotted versus thickness (50-500 nm) and wavelength (1100-2000 nm) in **Figure 4**. There is a transition region that develops with thicknesses less than 250 nm. The absorption becomes pinned at ENZ when crossover between transmission and reflection curves is fixed by a thick film.

With the desire to use thinner films, it is important to note that the best operating pump wavelength is not always at the ENZ wavelength. Although absorption spectra are broad for ENZ films (200-400 nm), for extremely thin films the best excitation regime is red-shifted.

Despite the fact that the absorption can shift for thinner films, the absorption curve is still broad and has weak concavity which is why the pump wavelength is not always crucial for ENZ films. However, for the probe wavelength shown in Figure 4b, the wavelength of interest is the point that has the most rapid change in refractive index. This occurs at ENZ regardless of the thickness of the film because the dispersion of the refractive index is strongest here. The selection of probe wavelength is thus more important than the pump for refractive index changes, although the ideal pump wavelength can shift more than 400 nm for films with thickness less than 200 nm. Because the free carrier process is directly related to absorption, this can effectively decrease the nonlinear IR efficiency by a percentage of absorption shift. For example, pumping a 100-nm film at ENZ would see a more than a 30% decrease in efficiency as opposed to pumping it at a red-shifted wavelength. For our films at ~200 nm, excitation at 1350, 1620, and 1890 nm have slightly different efficiencies (~5%) because the absorption is so broad for these thicknesses. This is reflected in the nonlinear calculations in the following discussion.

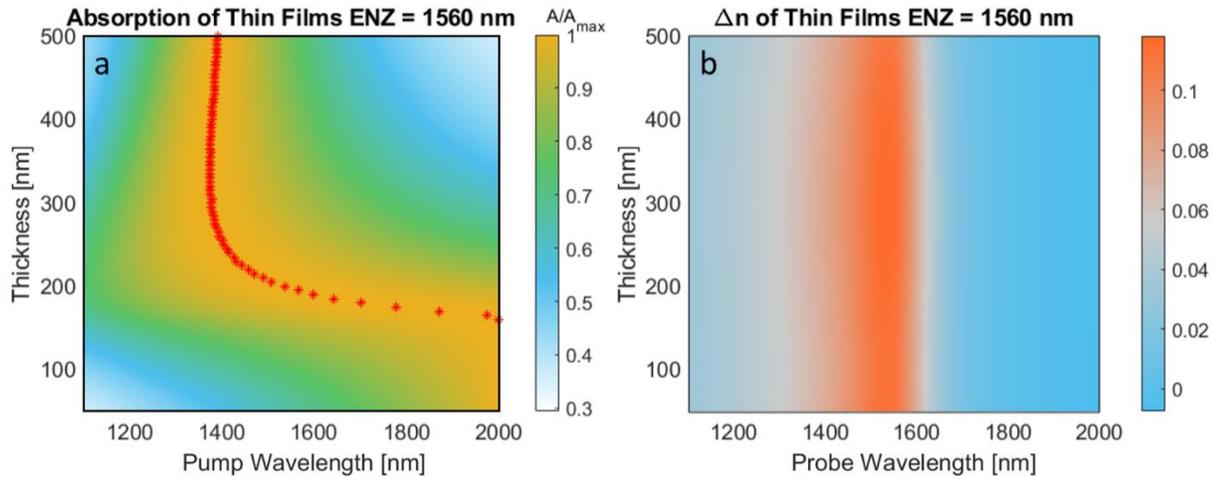

**Figure 4.** (a) Normalized absorption of thin films with ENZ wavelength at 1560 nm and varying thickness. Red stars represent the peak absorption for an individual film thickness. As the thickness of the film decreases, the absorption peak is broadened and red-shifted thus the optimal pumping wavelength for thin films is not always at ENZ. (b) Change in refractive index versus thickness and wavelength. The peak in Δn is strongest at the ENZ wavelength and thus the probe wavelength of choice should be at ENZ.

Considering all three films used in this study, we find that it would be helpful to the community if nonlinear coefficients for all the films are released. Typically, other papers in literature only report Δn values because the nonlinearities are not linear with irradiance, or only report Δn values for a single pump and probe wavelength combination. However, with our numerical model and our white light experimental technique, the changes in transmission and reflection experimentally measured for each film can be matched to theory and presented as $n_{2,\text{eff}}$ and $\beta_{\text{eff}}$ values. **Figure 5** and the supplemental material shows both $n_{2\text{eff}}$ and $\beta_{\text{eff}}$ for the three films pumped at 1350, 1620, and 1890 nm. It is important to note that the nonlinear coefficients are a function of excitation wavelength, angle, and excitation intensity ($n_{2,\text{eff}}(\lambda_{\text{pump}},\theta,I_{\text{pump}})$) and should be treated as such.

As shown in Figure 5 (a-c), the coefficients are relatively close in magnitude to each other for both sample and excitation wavelength, where $n_{2,\text{eff}} \sim 3.75\text{-}4.95\times10^{-3}$ cm$^2$GW$^{-1}$ and $\beta_{\text{eff}} \sim -360\text{-}450$ cmGW$^{-1}$. The intensity used is 30.5 GWcm$^{-2}$, at an angle of incidence normal to the sample. The external intensity incident on the film is calculated using the flat-top method. The film properties such as ENZ crossover point, thickness, and loss determine the location and spectral width of the coefficient curves. As an example, S4 has broader nonlinear spectrum compared to S1 and S5 because its overall ε'' is larger and the film is thinner compared to S5 with a similar $\lambda_{\text{ENZ}}$. Due to S4 having a larger ε'' it also has the largest $n_{2,\text{eff}}$ value because its base absorption of the pump is larger than S1 and S5. However, because all of the film thicknesses are ~200 nm, Figure 4a shows that the absorption curve is very broad and thus the difference between the 1350, 1620, and 1890 nm pump wavelengths is minimal. Finally, Figure 5d shows the nonlinear coefficients calculated at increasing intensities, where the response of electrons in the conduction band begin to saturate. This results in an overall decrease in the peak and a spectral red-shift.

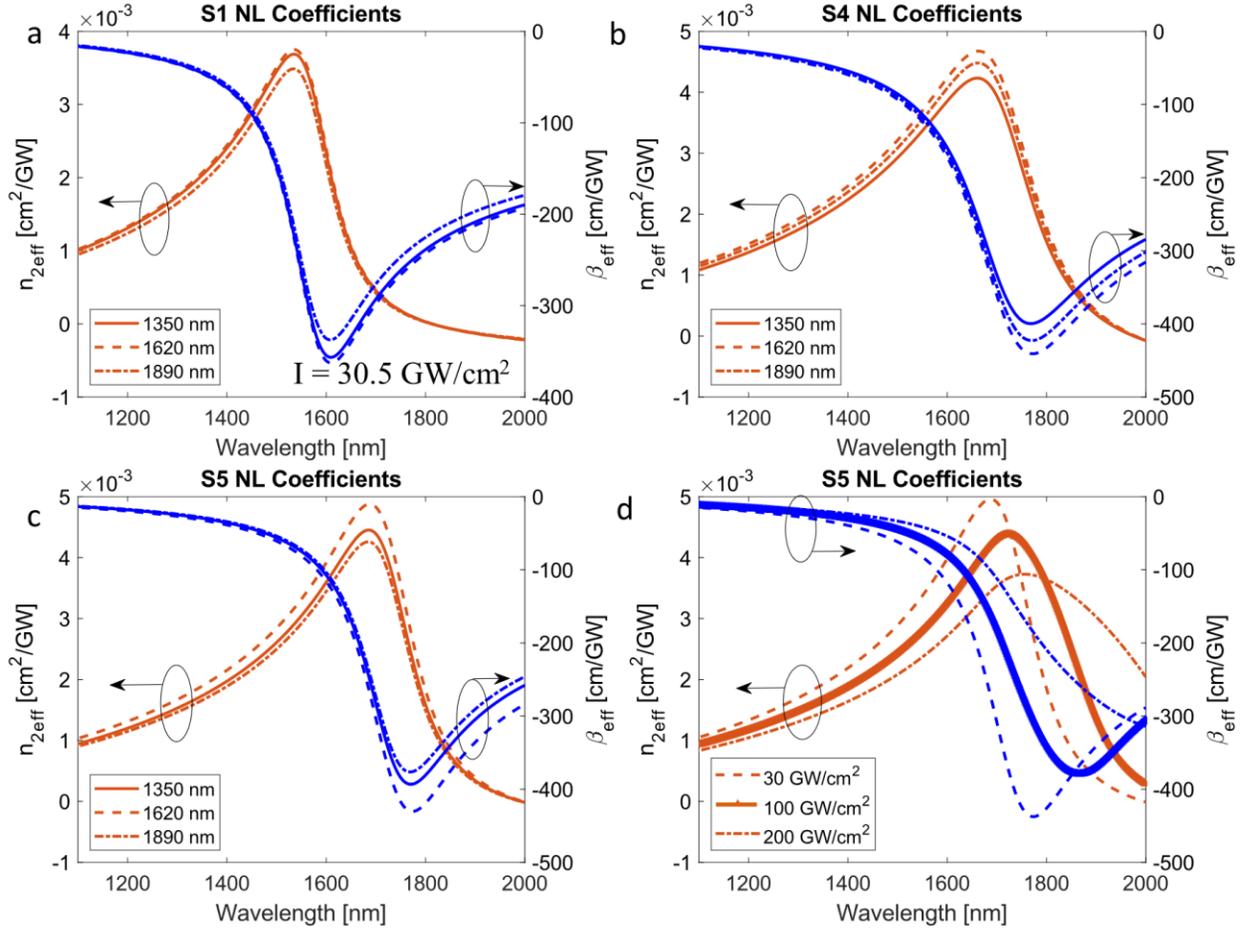

**Figure 5.** (a-c) S1, S4, and S5 nonlinear coefficients $n_{2,\text{eff}}$ and $\beta_{\text{eff}}$ versus wavelength extended to 1100-2000 nm. The intensity used is 30.5 GWcm$^{-2}$ at an angle normal to the sample. (d) S5 nonlinear coefficients dependent on intensity, showing saturation and red-shifting of the nonlinear coefficients with increasing intensity.

## 5. Conclusion

We report for the first time nonlinear optical measurements of GZO films with varying ENZ crossover points. The films are studied and modeled according to their nonlinear optical reflection and transmission properties when excited with infrared pump beams. Using a white light probe, both wavelength and temporal data were obtained and numerically modeled. Typical nonlinear experiments only produce nonlinear coefficients for single pump and probe wavelengths. Here, nonlinear coefficients $n_{2\text{eff}}$ and $\beta_{\text{eff}}$ are calculated for all the films for all spectral regions of the probe while discussing optimal probe and excitation wavelengths for thin films. Additionally, all three of the film's data are released and available to be used and modeled for the community in the supplemental section. For a while, nonlinear researchers have looked for new, flashy materials. ENZ materials are now targeted to be used as tools; implemented into devices and systems aiming for high modulation with relatively low activation requirements.

**Experimental Setup**

The experimental setup for the reflection and transmission measurements is shown in Figure 1 (a). Pump-probe measurements of transient transmission and reflectivity were performed by splitting the fundamental 800 nm output of a 5 kHz, 60 fs Ti:sapphire laser amplifier (SpectraPhysics) into two branches. The pump branch of the laser was directed through an optical parametric amplifier to generate pulses of variable wavelengths. Afterward, this beam was chopped mechanically to 2.5 kHz. The probe branch was focused through a 5 mm thick sapphire crystal to generate a supercontinuum white light, which was filtered for the near-infrared ($\lambda_{probe}$= 1000-1600 nm) and polarized for cross-polarization with respect to the pump beam to avoid two-beam coupling. [54,55] This does not affect the nonlinearity because free carrier effects are isotropic.[35] The pump and probe were spatially overlapped at the sample position with an average pump sample spot size of 700 µm diameter (defined as 1/e of the peak intensity). Both transmitted and reflected probe light were collected and directed to a pair of CCDs (Ultrafast Systems Cam NIR-2). The transient signal was calculated from the pump-on versus pump-off difference using the mechanical chopper, thus generating differential reflection $\Delta R/R$ and transmission $\Delta T/T$ signals directly. The transmission and reflection data are measured at 5° with respect to the sample normal.

**Acknowledgements**

A.B., R.S., N.K. acknowledge the Air Force Office of Scientific Research (Nos. FA9550-1-18-0151 and FA9550-16-10362).

Work performed at the Center for Nanoscale Materials, a U.S. Department of Energy Office of Science User Facility, was supported by the U.S. Department of Energy, Office of Basic Energy Sciences under Contract No. DE-AC02-06CH11357.

**Data Availability**

The data that support the findings of this study are available in the supplementary material of this article and at https://nanohub.org/tools/nlenz/.

**Conflict of Interest**

The authors have no conflicts to disclose.

Table of Contents

**Gallium-doped Zinc Oxide: Nonlinear Reflection and Transmission Measurements and Modeling in the ENZ Region**

*Adam Ball[1*], Ray Secondo[1,2], Benjamin Diroll[3], Dhruv Fomra[1], Vitaly Avrutin[1], Ümit Özgür[1], and Nathaniel Kinsey[1]*

Here we explore the nonlinearities in Gallium-doped Zinc Oxide in the epsilon near zero (ENZ) region and show enhanced $n_{2,\text{eff}}$ values on the order of $4.5\times10^{-3}$ $cm^2GW^{-1}$ for three different IR pumping wavelengths on 200-300 nm thin films. Using a numerical model, we discuss optimal pumping and probing wavelengths for ENZ films.

ToC Figure

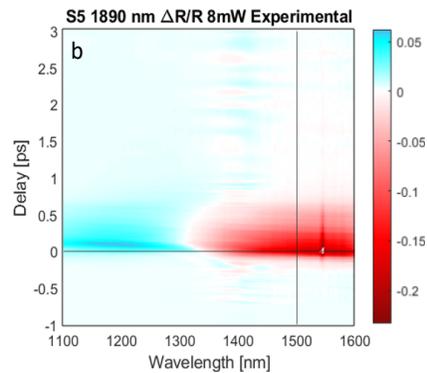

Supporting Information

**Gallium-doped Zinc Oxide: Nonlinear Reflection and Transmission Measurements Supplemental Information**

*Adam Ball[1*], Ray Secondo[1,2], Benjamin T. Diroll[3], Dhruv Fomra[1], Vitaly Avrutin[1], Ümit Özgür[1], Nathaniel Kinsey[1]*

1 Department of Electrical and Computer Engineering, Virginia Commonwealth University, Richmond, VA, 23284, USA
2 Air Force Research Laboratory Wright-Patterson Air Force Base Dayton, OH, 45433, USA
3 Center for Nanoscale Materials, Argonne National Laboratory, 9700 S. Cass Avenue, Lemont, IL, 60439, USA

**S1 Material Growth & Linear Optical Properties**

The growth of the ZnO layers was carried out by plasma-enhanced molecular-beam epitaxy. Knudsen cells were used to evaporate Zn, Ga, and Mg, and a radio frequency plasma source operating at 400 W was employed for the reactive oxygen source. For growths directly on c-sapphire, the substrate was thermally cleaned at 730 °C for 20 min under oxygen plasma exposure. Then, 5-nm MgO buffer was deposited at substrate temperatures, $T_S$ = 730 °C, followed by the growth of 10-nm low-temperature-ZnO buffer at $T_S$ = 280 °C. After annealing of the buffer layer stack at 750 °C for 5 min, a 15-nm-thick high-temperature-ZnO layer was deposited at $T_S$= 680 °C under oxygen-rich conditions. The 200-nm Ga-doped ZnO (GZO) layer was grown at 350 °C under metal-rich conditions. Electron concentration and, thus, crossover frequency in the grown structures was tailored by varying Ga flux by changing the temperature of the Ga Knudsen cell. The growth progression was monitored in situ using reflection high-energy electron diffraction (RHEED).

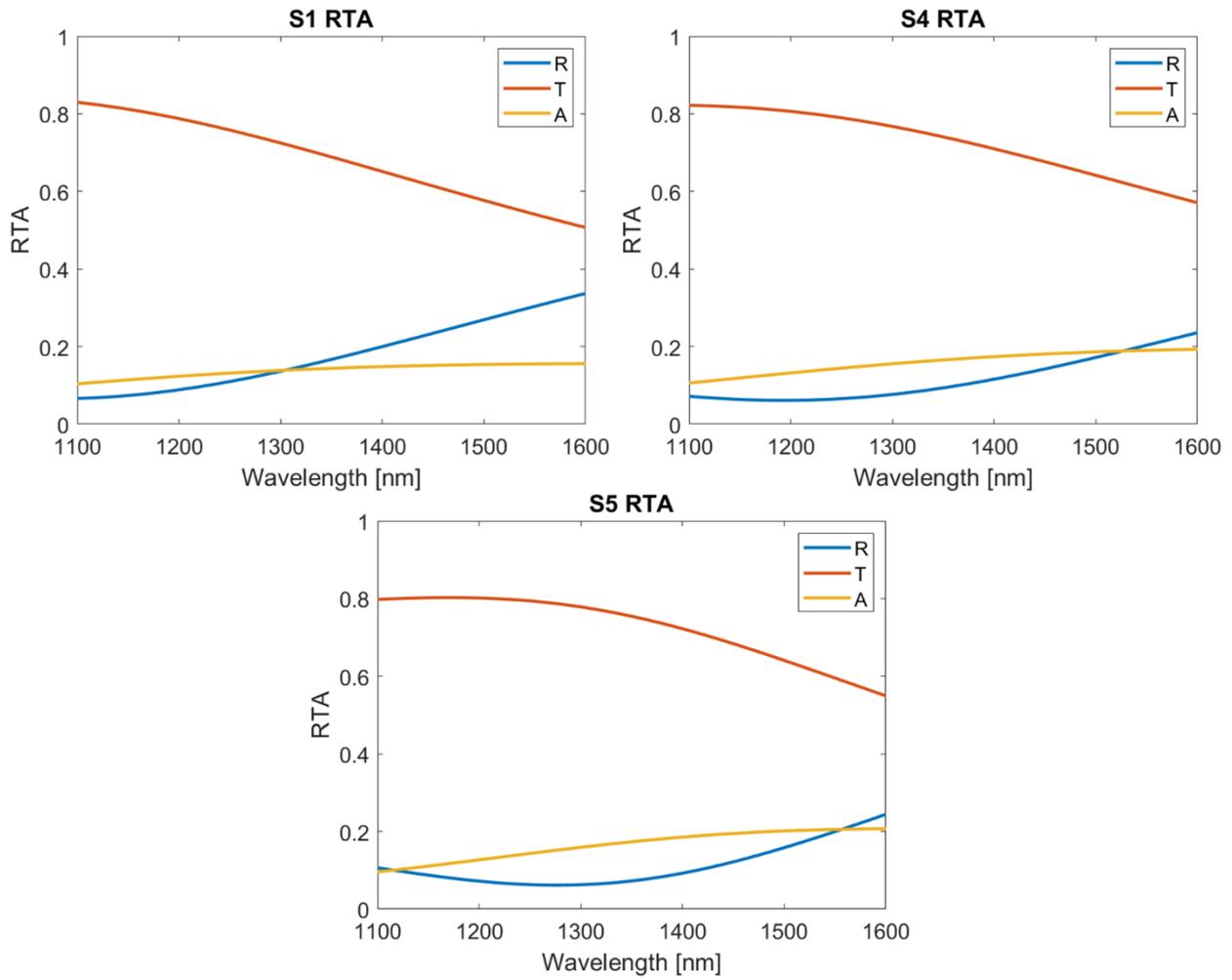

**Figure A1.** Sample reflection, transmission, and absorption calculated from transfer-matrix-method as mentioned in section 2 of the main text.

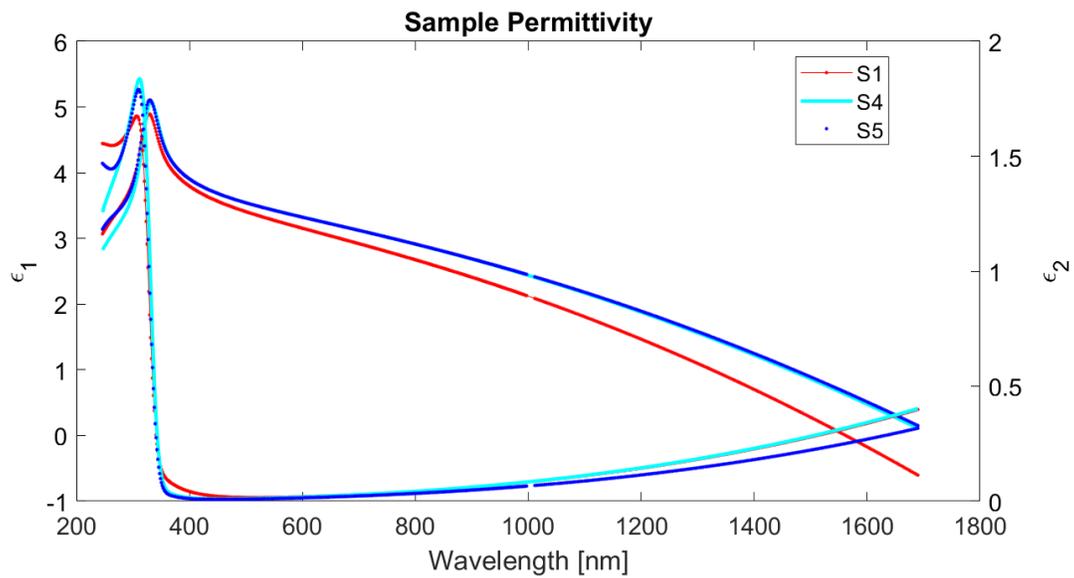

**Figure A2.** Sample permittivity of all three GZO samples extracted from J. A. Woollam M-2000 Ellipsometer. Shows the full dispersion of the films including the bandgap that occurs at approximately 350 nm for all six samples. Where Figure 1b in the paper is zoomed in on the region of interest.

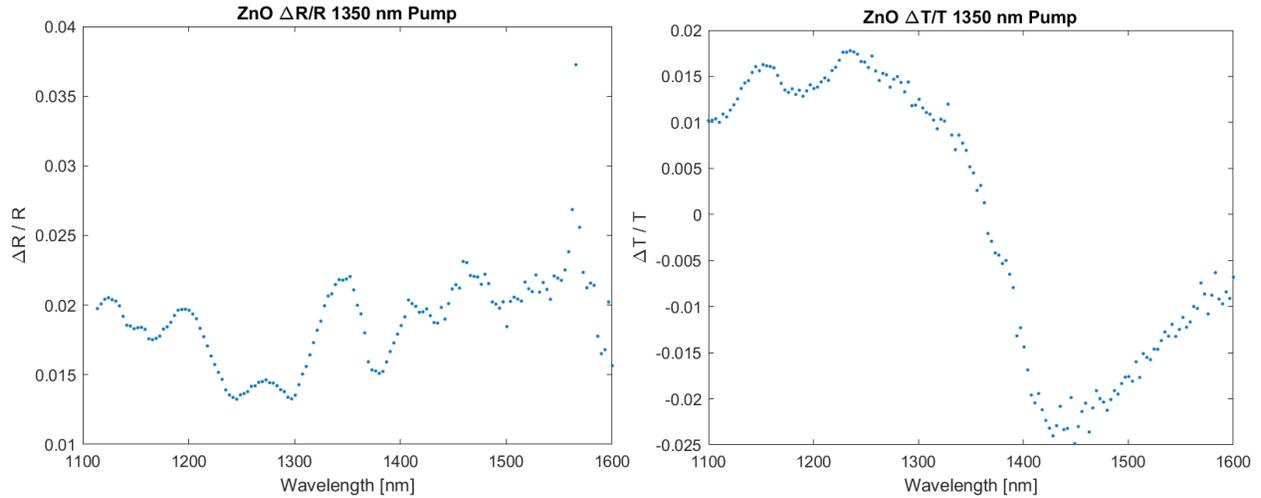

**Figure A3.** Experimental data on substrate composed of ZnO/MgO/Sapphire showing the magnitude at the highest intensity is 2% $\Delta R/R$. When normalized by linear $R$ this contribution becomes within experimental error.

## S2 Data Filtering

Because the probe beam is from a white-light source coupled to a camera, chirp shows up in the raw data signals. This often means the shorter wavelength ranges have the nonlinear transient occur before the longer wavelengths. Using Ultrafast Systems' *Surface Xplorer* software, the signal could be chirp corrected so that the nonlinear signal occurs at $t = 0$. The files and data attached in the supplemental section are chirp corrected.

The pump and probe beams were cross-polarized to reduce the effect of two-beam coupling and filters were used to minimize the pump contribution into the detectors. However, some amount of the pump beam scattered into the detector in various measurements because of degeneracy conditions (namely the 1350 nm and 1620 nm IR pump experiments). This contribution is constant throughout all times of the signal and is evident in the detector signal centered around the pump wavelength. To remove this effect, the data was averaged and subtracted at a time (t ~ -6 ps) before the nonlinear transient (t ~ 0 ps). This post-treatment of the data, done in the *Matlab* code provided, resolves the wanted transients while removing the noise. Furthermore, two of the pixels on the CCD are faulty and cause a spike in the signal around 1560 and 1600 nm. This is evident in the dispersion data and is recognized but ignored.

For IR excitation conditions, the samples were exposed substrate-side first, where the skin depth at the IR pump/probe wavelengths is greater than the thickness of the film. A ZnO/MgO/Sapphire film was measured at all three pumping wavelengths to explore the substrate contribution. In all cases, the contribution of the substrate is <5% of the nonlinear effects from the GZO film (within experimental error). This data is presented in supplemental A3. The nonlinear contribution from the substrate is within experimental error of other parameters and thus ignored.

**S3 Data Release**

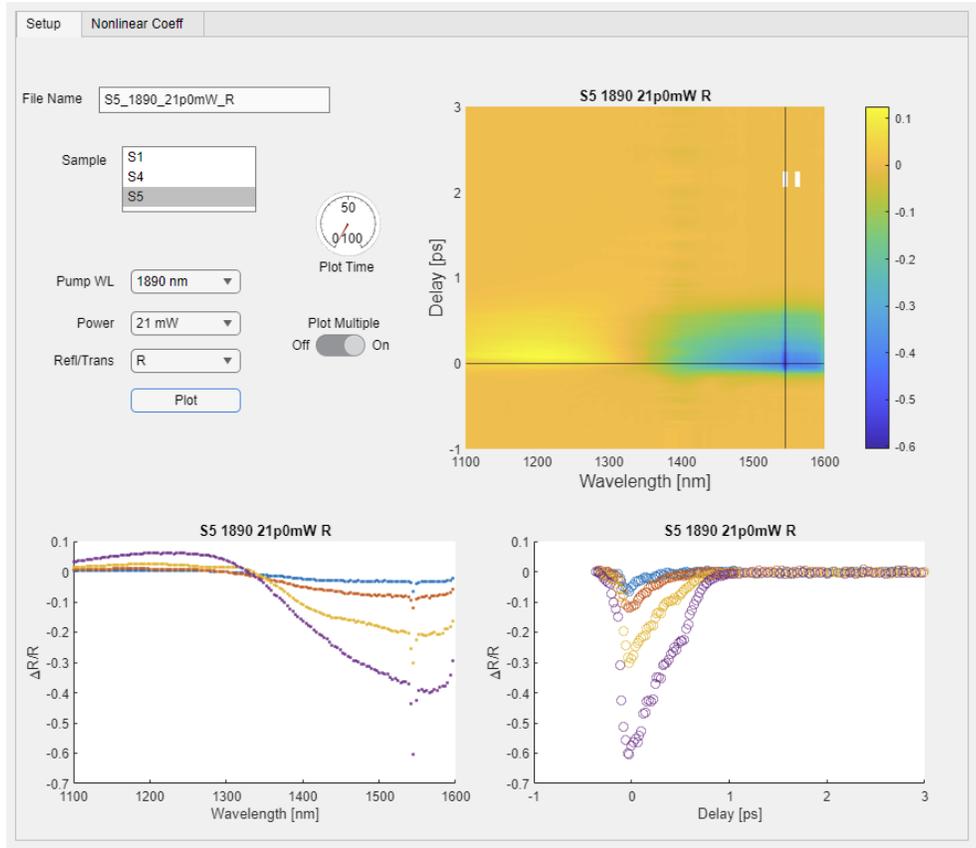

**Figure A4**: Nonlinear GZO measurements app. This Matlab app holds all the IR data used in this study for all three samples and can show fits and includes the $n_{2\text{eff}}$ and $\beta_{\text{eff}}$ coefficients for each film.

All three films measured at all three excitation wavelengths and all five powers are available for access online. This rich set of data is a part of the NLENZ app posted on https://nanohub.org/tools/nlenz/. The data is included on the "Example Data" tab which the models are already incorporated. The data can be fully exported and used in simulation elsewhere. In addition, the app includes the ability to model/predict the film properties under other excitation, wavelength, or temporal conditions.